# Reversal of strain state in a Mott insulator thin film by controlling substrate morphology

Reetendra Singh[1], Abhishek Rakshit[1], Galit Atiya[1], Michael Kalina[1], Yaron Kauffmann[1], and Yoav Kalcheim[1]*

[1]Department of Materials Science and Engineering, Technion-Israel Institute of Technology, Haifa 32000, Israel

*Email: ykalcheim@technion.ac.il

**ABSTRACT**

The $V_2O_3$ phase diagram contains two insulating phases and one metallic phase with different lattice structures. The stability of these phases is very sensitive to pressure, offering a mechanism to tune phase transitions by inducing strain in thin films. The most studied source of strain is lattice mismatch between the film and the substrate. In this work, however, we find that the film/substrate thermal expansion mismatch can be made to play a dominant role by modifying the substrate morphology. When grown on sapphire, the lattice mismatch induces compressive strain in the $V_2O_3$ films, whereas thermal expansion mismatch induces tensile strain. We find that minute changes in substrate morphology may relax the compressive strain component, allowing the thermally-induced tensile component to overcome it. Thus, by simple annealing of the substrates to create either a flat or stepped morphology, strongly compressive or tensile strains may be induced in the films. This results in either full suppression of the metal-insulator transition or stabilization of insulating phases at all temperatures, exhibiting many orders of magnitude differences in film resistivity. To elucidate the strain relaxation mechanism, we use high-resolution scanning transmission electron microscopy (HRSTEM) to image the atomic steps in the substrate and the adjacent crystallographic defects in the $V_2O_3$. These findings offer a hitherto underexplored mechanism to tune strain in thin films, deepen our understanding of the effects of structural degrees of freedom on phase stability of a canonical Mott insulator and may allow for applications requiring insulator-metal switching above room temperature.

## INTRODUCTION

V$_2$O$_3$ is a prototypical Mott system and one of the most widely studied quantum materials, owing to its rich phase diagram and a first-order metal-insulator transition (MIT) that involves a strong interplay of structural, electronic, magnetic, and orbital degrees of freedom [1–4]. At room temperature and ambient pressure, bulk V$_2$O$_3$ is a paramagnetic metal (PM) with a corundum structure. When cooled it undergoes a transition to a monoclinic antiferromagnetic insulator (AFI) at a transition temperature of ~160 K [4]. In bulk V$_2$O$_3$, the resistivity changes across the MIT by more than seven orders of magnitude [1]. While the pronounced sensitivity of V$_2$O$_3$ and other strongly correlated electron systems to external stimuli underpins their potential for applications in memory devices, resistive RAM selectors, optical switches, and neuromorphic systems, their MIT behaviour is not yet fully understood [5–9]. Previous reports suggested that the properties of Mott insulators films may be influenced by growth temperature, pressure of oxygen partial pressure, substrate orientations and its treatment [10–16]. Strain is a key factor influencing the MIT in V$_2$O$_3$ films, as it directly modifies orbital overlap and lattice distortions that govern the electronic and magnetic states [17–27]. Both defects and substrate-induced strain play a decisive role in tuning the transition [22–27], highlighting the importance of strain engineering in controlling MIT behavior. Pofelski *et al.* demonstrated that V$_2$O$_3$ films of varying thicknesses grown on c-sapphire exhibit a strongly suppressed MIT, with the extent of suppression depending on the film thickness [28]. Additionally, a wide range of strain control in V$_2$O$_3$ and (V$_{0.985}$Cr$_{0.015}$)$_2$O$_3$ films on *c*-sapphire was realized by inserting (Cr$_{1-\gamma}$Fe$_\gamma$)$_2$O$_3$/Cr$_2$O$_3$ buffer layers and systematically varying the Fe content [29].

Despite this large body of work, several crucial aspects of strain engineering in V$_2$O$_3$ are not well understood - namely the interplay between different mechanisms for strain formation such as lattice mismatch, thermal expansion mismatch and strain relaxation by defect formation. Indeed, it is often found that the characteristics of the transition such as the critical temperature, thermal transition width and hysteresis vary widely depending on the choice of substrate, but also for seemingly similar substrates. To understand the complete phenomenology observed in V$_2$O$_3$ thin films we focus on sapphire substrates, which are the most widely used since the two materials are isostructural, allowing for heteroepitaxial growth [10–12,28–32]. The lattice constants of sapphire are smaller than those of V$_2$O$_3$ - by 7.8% in the *c*-axis and 4.1% perpendicular to it. This may result in large compressive epitaxial strain in the V$_2$O$_3$ films whose magnitude depends on strain relaxation mechanisms taking place during growth. An additional source of strain is the substrate/film thermal expansion mismatch. Comparing lattice constants of V$_2$O$_3$ and sapphire between typical growth temperatures (~1000 K) to room temperature, one finds differential lattice constant mismatch between the substrate and film which would result in ~0.9% compressive strain in the *c*-axis and ~0.74% tensile perpendicular to the *c*-axis (assuming perfect clamping) [33,34]. Therefore, when V$_2$O$_3$ films are grown on *c*-plane oriented sapphire the lattice constant mismatch tends to compress the films in the plane while the thermal expansion mismatch has an opposite effect. Due to the high sensitivity of Mott insulators to strain, understanding the role of these competing effects is of paramount importance. Here we show that by controlling the substrate morphology, the degree of strain relaxation at the growth temperature may be controlled, thereby diminishing the compressive strain component and driving the film into a tensile strain regime upon cooling. It is found that a simple annealing pretreatment of the substrate allows to tune the properties of V$_2$O$_3$ from a state of nearly complete MIT suppression and metallic behaviour at low temperatures to a fully developed resistive transition and a paramagnetic insulating (PI) phase at room temperature. By imaging the defects which form in the $V_2O_3$ films with high resolution STEM and correlating them with substrate morphology, we shed light on the mechanism by which strain is induced in $V_2O_3$ films. This method of strain engineering can be used to drive $V_2O_3$ into different regimes of the phase diagram, induce an insulating state above room temperature, and vary resistance in different films by more than 7 orders of magnitude. The strain control mechanism, which is elucidated here, may be applicable to many other systems where epitaxial and thermal strains play a significant role.

**RESULTS**

100 nm heteroepitaxial thin films of $V_2O_3$ were grown on sapphire substrates with different orientations, using RF magnetron sputtering from a stoichiometric $V_2O_3$ target. The films were grown in a 5 mTorr Ar atmosphere at a substrate temperature of 750 °C, followed by thermal quenching at a rate of ~90 °C/min [28]. Resistance versus temperature measurements reveal that the film grown on (0001)-sapphire exhibits a strongly suppressed MIT, in contrast to films grown on other sapphire orientations, which show a pronounced MIT with a resistance change of 6-7 orders of magnitude **(Supplementary figure 1)** [21]. In the films grown on *c*-sapphire, the in-plane lattice is aligned parallel to the substrate surface and thus constrained by it **(Supplementary figure 2)** [26,28]. The confinement restricts the lattice expansion needed for the corundum-to-monoclinic phase transition, which in turn suppresses the MIT. In this study we examined the effect of substrate morphology on the strain induced in films grown on (0001)-sapphire, the mechanism by which the substrate morphology affects the strain and the resulting film transport properties.

To control the substrate morphology, c-sapphire was annealed at 1000 °C, 1200 °C, and 1400 °C for 12 hours in presence of air. A detailed surface characterization of both as-received and annealed sapphire substrates was performed using atomic force microscopy (AFM) to evaluate the surface morphology (see **Figure 1**). **Figure 1(a)** shows the AFM image of the surface morphology for the as-received sapphire substrate, **Figure 1(b)** for the sapphire substrate after film growth but without annealing (measured on parts of the sapphire substrates where no deposition occurred due to clamp shadowing), and **Figure 1(c-e)** for annealed sapphire under different conditions. The as-received sapphire substrates exhibit high surface roughness, while thermal annealing significantly smooths the surface, leading to low-roughness thermally equilibrated step-terraced surfaces [35]. Previous reports suggest that the dimensions of atomic steps depend on both the annealing conditions and the impurity level of the sapphire substrate [35–38]. **Figure 1(f-j)** shows the height profile of the atomic steps. In our case, annealing at 1000-1400 °C results in increasing step heights and width as a function of temperature. The root-mean-square roughness ($Rq$) of the as-received sapphire substrate was found to be ~0.6 Å. Already at the film growth conditions small atomic steps are formed, although they are not as wide and straight as those observed during the dedicated annealing process **(Figure 1(b))**. This sample has the lowest roughness (0.1 Å) and increases monotonously up to ~1.0 Å for the highest annealing temperature.

100 nm thin films of $V_2O_3$ were then grown on annealed or as-received sapphire substrates using RF magnetron sputtering. $\phi$-scans for the $(20\bar{2}2)$ peak of $V_2O_3$ showed three peaks at intervals of 120°, following the crystallographic orientation of the sapphire, revealing an epitaxial relationship between the films and substrates (**Supplementary figure 3** and see also [28]). *θ-2θ* X-ray diffraction (XRD) scans reveal only the (000*l*) peaks of $V_2O_3$, alongside the sapphire peaks **(Supplementary figure 4(a))**. X-ray reflectivity (XRR) measurements **(Supplementary figure 4(b, c))** performed on $V_2O_3$ films grown on all sapphire substrates confirm that all films have identical thicknesses and densities up to the measurement resolution, further supporting the consistency in film quality.

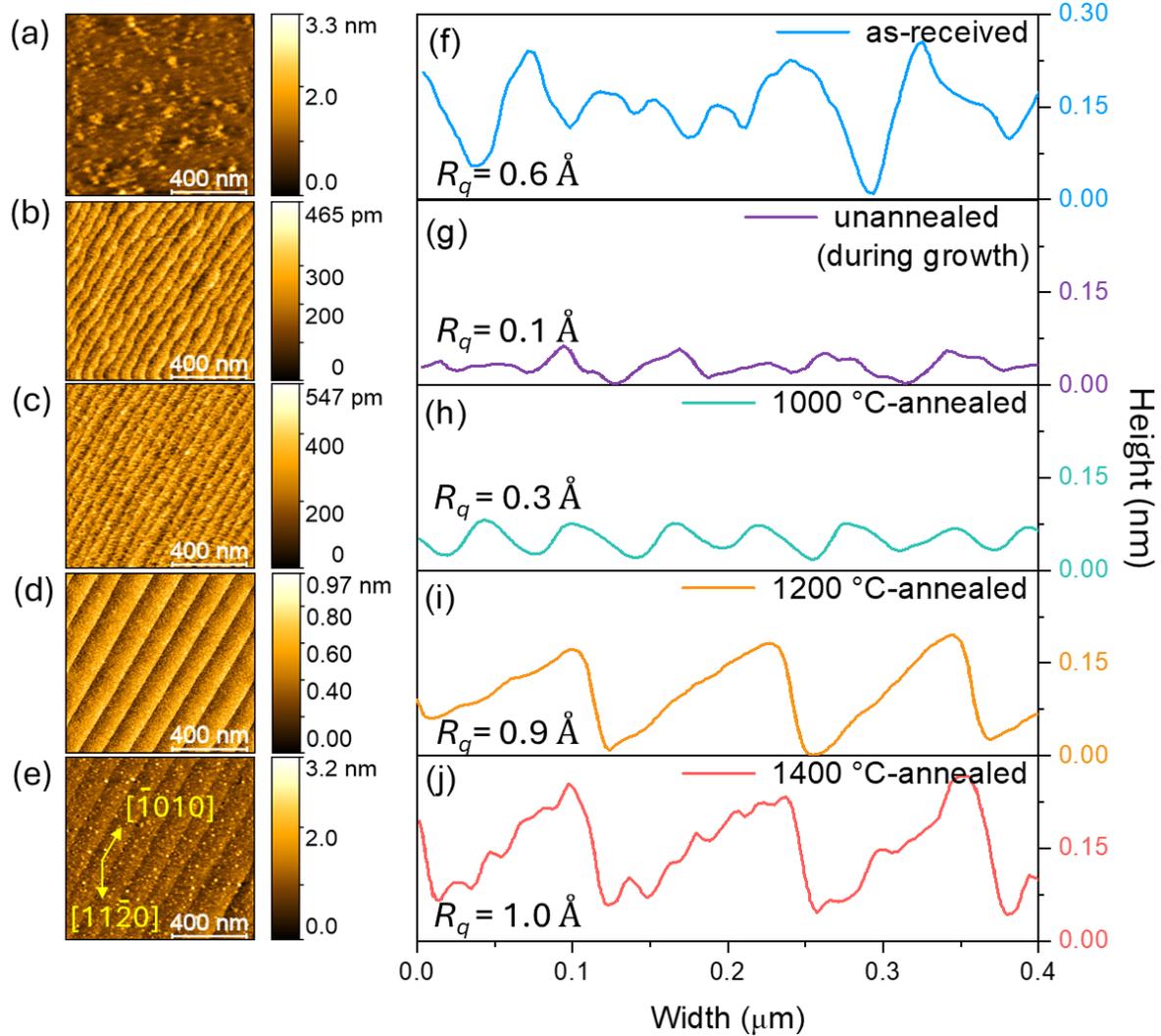

**Figure 1.** *Effect of annealing on sapphire substrate morphology: (a-e) AFM images of $V_2O_3$ thin films of as-received c-plane sapphire showing no atomic steps, whereas annealed substrates exhibit well-defined atomic steps. (f-j) height profile of the steps taken from the topographic image; the corresponding root mean square roughness ($R_q$) values are indicated. The step heights and terrace widths can be tuned by varying annealing conditions.*

Lattice parameters and strain in the films were evaluated by 2θ/ω XRD scans on the $(20\bar{2}2)$ and (0006) Bragg reflections shown in **Figure 2 and Table 1.** The $V_2O_3$ film grown on unannealed sapphire exhibits -0.54±0.06% compressive strain in the *ab*-plane and +0.48±0.03% tensile strain along the *c*-axis (**Figure 2(b) and 2(a)**) with respect to bulk $V_2O_3$. The asymmetric peak profile with a tail towards lower 2θ in the (0006) reflection indicates the presence of strain gradients along the *c*-axis. With increasing substrate annealing temperature, a clear shift of the (0006) reflection peaks toward higher 2θ values are observed, surpassing the bulk value, signifying a transition from tensile to compressive strain in the *c*-axis. **(Figure**

**2(a))**. Consistent with the Poisson effect the $(20\bar{2}2)$ reflection shifts from compressive to tensile strain as the substrate annealing temperature rises. The peak positions for the film grown on sapphire annealed at 1400°C is closer to that of the corresponding PI phase [29,39,40]. Additionally, with increasing substrate annealing temperature, the peaks become more symmetric, indicating a diminishing strain gradient within the films.

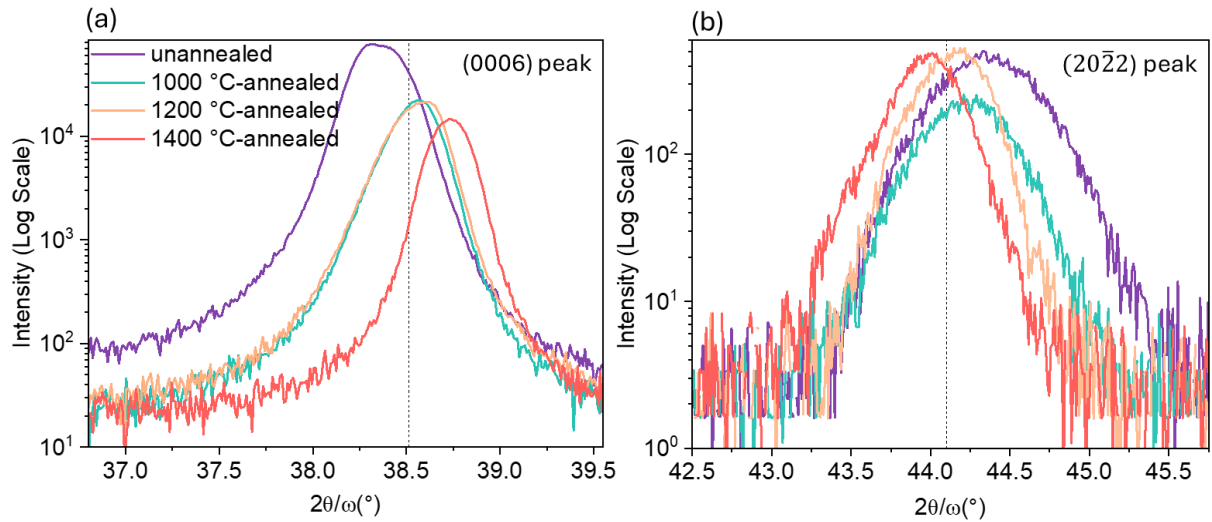

**Figure 2. *Reversal of strain regimes for $V_2O_3$ films grown on sapphire substrates with different annealing parameters*:** *(a) Out-of-plane $2\theta/\omega$ scan of the (0006) XRD peak for $V_2O_3$ films grown on sapphire substrates shows a systematic shift toward higher angles with increasing annealing temperature of the sapphire, indicating a transition from tensile to compressive strain consistent with out-of-plane lattice contraction. (b) The corresponding $2\theta/\omega$ scan of the $(20\bar{2}2)$ peak shifts toward lower angles, further confirming strain relaxation in the $V_2O_3$ films due to annealing-induced surface modifications of the sapphire. The dotted lines indicate the peak positions corresponding to bulk $V_2O_3$.*

| annealing conditions | $a$ [Å] | in-plane strain $\varepsilon_a$ (%) | $c$ [Å] | out-of-plane strain $\varepsilon_c$ (%) |
|---|---|---|---|---|
| unannealed | 4.924 ± 0.003 | −0.54±0.06 | 14.070 ± 0.004 | +0.48±0.03 |
| 1000 °C-annealed | 4.941 ± 0.003 | −0.20±0.06 | 14.001 ± 0.004 | −0.01±0.03 |
| 1200 °C-annealed | 4.951 ± 0.003 | 0.00±0.06 | 13.996 ± 0.004 | −0.05±0.02 |
| 1400 °C-annealed | 4.972 ± 0.003 | +0.43±0.06 | 13.938 ± 0.004 | −0.47±0.02 |

**Table 1.** *Annealing conditions of sapphire, lattice parameters, and the corresponding in-plane and out-of-plane strain within the films.*

In reciprocal space mapping (RSM) of the symmetric (0006) reflection, clear differences are observed between $V_2O_3$ films grown on unannealed and annealed *c*-sapphire substrates **(Supplementary figure 5)**. For films on unannealed sapphire, the (0006) peak is broad along $Q_z$ and sharp along $Q_x$, indicating significant strain gradients along the growth direction, while the in-plane crystalline alignment remains relatively uniform. In contrast, films grown on annealed sapphire exhibit the opposite behaviour, a sharp (0006) peak in $Q_z$ and broadening in $Q_x$, signifying relaxation in epitaxial alignment but the presence of in-plane mosaic tilt due to defects, likely arising from step-influenced growth or anisotropic relaxation at terraced interface.

To evaluate local strain in the films and understand its relationship to the substrate morphology, high-resolution scanning transmission electron microscopy (HRSTEM) was performed. Since the length of the steps appeared close to the $[\bar{1}010]$ crystallographic direction, lamellae were cut perpendicular to this orientation so that steps were readily visualized in the lamella. **Figure 3(a)** shows HAADF-STEM images of the substrate/film interface for a $V_2O_3$ film grown on a sapphire substrate annealed at 1200 °C. Two adjacent atomic steps are observed in the substrate (see dotted arrows). To visualize the effect of these steps on the $V_2O_3$ film, geometrical phase analysis (GPA) from STEM image; the deformation of the lattice parameters of the $V_2O_3$ with respect to the sapphire [41,42] was carried out near the vicinity of steps using two g-vectors corresponding to the planes $g_1=(0001)$ and $g_2=(\bar{1}2\bar{1}0)$ (out-of-plane and in-plane respectively- see **Supplementary figure 6(b)**). The (0001) Bragg-filtered image in **Figure 3(b,d)** reveals dislocations specifically localized near the atomic steps, while the $(\bar{1}2\bar{1}0)$ Bragg-filtered image **(Figure 3(c))** highlights stacking fault formed between the two adjacent steps. In-plane component of the strain tensor ($\varepsilon_{xx}$) obtained by GPA shows that the $V_2O_3$ film maintains some degree of epitaxial coherence with the substrate up to the stacking fault **(Figure 3(e))**, but exhibits a marked reduction in strain beyond it, indicating the onset of relaxation.

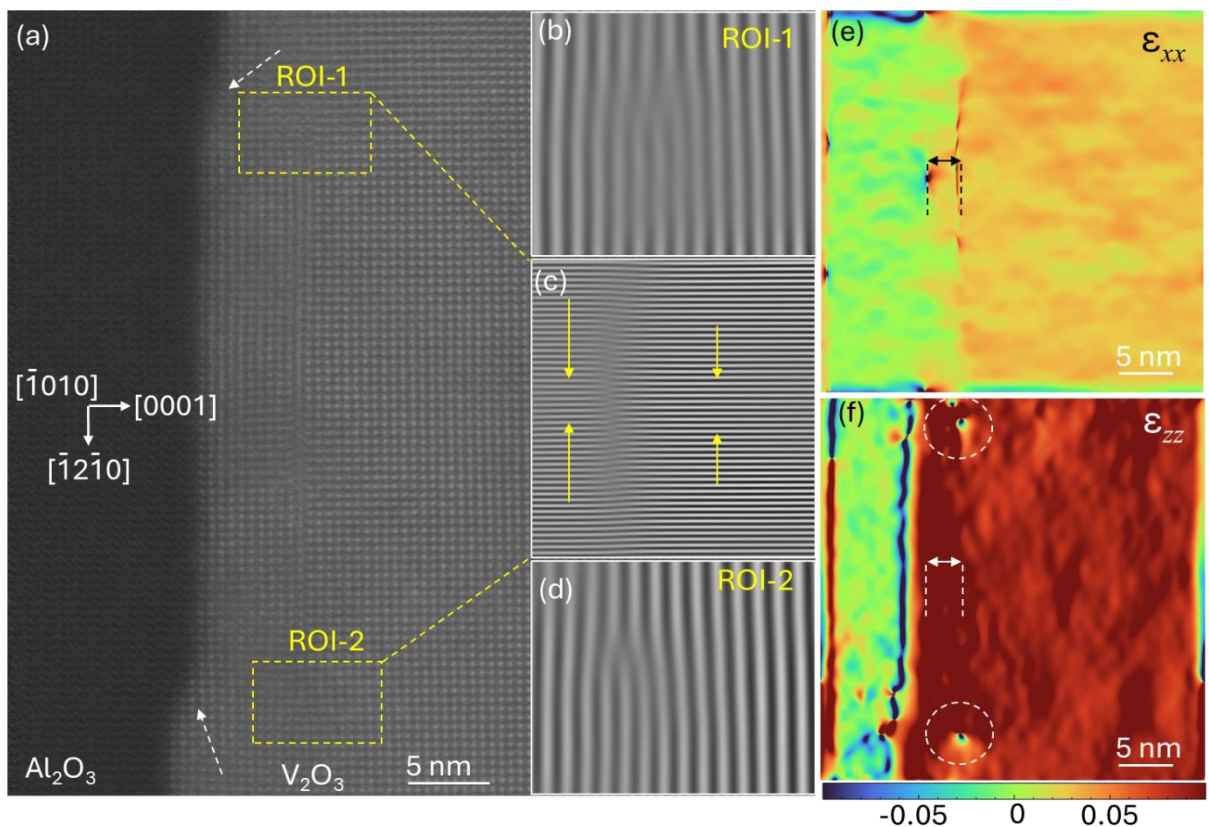

**Figure 3.** *Imaging of dislocations and stacking faults in the vicinity of substrate steps and their effect on strain relaxation: (a) HRSTEM image showing two adjacent atomic steps on sapphire annealed at 1200 °C. (b, d) The (0001) Bragg-filtered image of region of interest (ROI) generated using geometric phase analysis (GPA) confirms the presence of dislocations in the vicinity of the steps. (c) Bragg-filtered $(\bar{1}2\bar{1}0)$ image confirms the presence of a stacking fault between the steps/dislocations. The stacking fault serves as a relaxation mechanism for*

*the strong lattice mismatch-induced compressive strain adjacent to the substrate. (e) In-plane component of the strain tensor ($\varepsilon_{xx}$) reveals epitaxial coherence of the first ~2 nm of the $V_2O_3$ film, with significant strain relaxation observed beyond the dislocations and the intermediary stacking fault. (f) Out-of-plane component of the strain tensor ($\varepsilon_{zz}$) shows tensile strain near the substrate, up to the dislocations and stacking faults. These observations confirm that the substrate atomic steps promote strain relaxation by facilitating the formation of dislocations and stacking faults.*

Out-of-plane component of the strain tensor ($\varepsilon_{zz}$) shows an opposite trend due to the Poisson effect: a large tensile strain is observed between the substrate and the stacking fault and relaxes beyond it **(Figure 3(f))**. These results demonstrate that the substrate atomic steps facilitate the formation of dislocations and stacking faults, promoting strain relaxation during film growth.

To further investigate the role of atomic steps in other films, HAADF-STEM was performed on samples grown on sapphire substrates which were annealed at different temperatures. **Figure 4(a-d)** shows lattice-resolved images of $V_2O_3$ and $Al_2O_3$, confirming a flat interface for films grown on unannealed sapphire, whereas atomic steps appear on the surface of annealed sapphire (highlighted by dotted arrows). These steps have heights corresponding to sub-unit-cell dimensions. The effect and size of these atomic steps increase with annealing temperature of sapphire, as notably observed for the 1200 °C and 1400 °C annealed substrate compared to 1000 °C. Inset of **Figure 4(a-d)** presents (0001) Bragg-filtered images near the steps, which reveals localized defects concentrated near the atomic steps of the annealed sapphire and their absence in unannealed sapphire. The interface of the film grown on 1400 °C annealed sapphire appears significantly more complex, suggesting enhanced defect formation. **Figure 5(e-h)** confirms that associated defects facilitate strain relaxation in $V_2O_3$ films on annealed sapphire. These microscopy results confirm the early relaxation of $V_2O_3$ in presence of atomic steps of sapphire. Previous reports also suggested $V_2O_3$ films grown on annealed sapphire relaxes after approximately ~2-4 nm of growth [11,13].

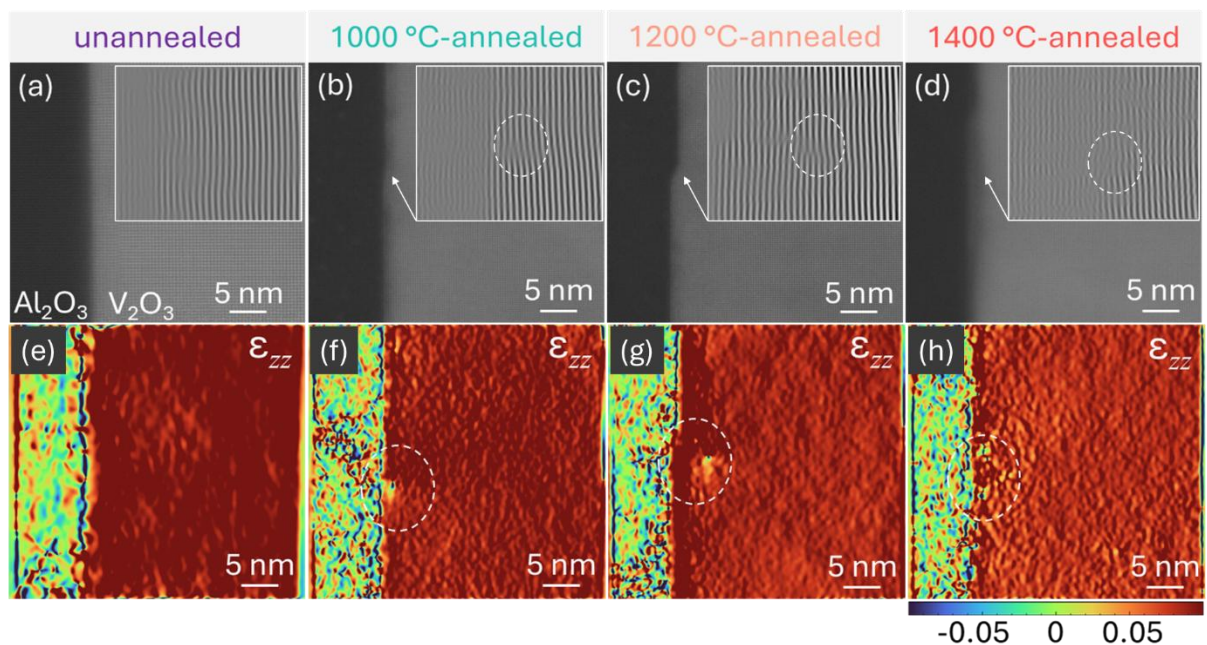

**Figure 4.** *Effect of substrate annealing at different temperatures on strain relaxation through formation of dislocations in $V_2O_3$ films: (a-d) HRSTEM images showing atomic steps*

*on the surface of annealed c-plane sapphire, which are absent in unannealed sapphire. The step heights correspond to sub-unit cell dimensions. Inset: (0001) Bragg-filtered image highlighting localized defects near the atomic steps in the sapphire substrate. (e-h) Out-of-plane strain maps reveal high tensile strain in the film on unannealed sapphire, and strain relaxation occurs in films on annealed sapphire, particularly near the dislocations. These sub-unit-cell steps and associated defects promote strain relaxation in $V_2O_3$ films grown on c-plane sapphire, facilitating partial accommodation of lattice mismatch and reducing in-plane strain.*

Subtle changes in substrate morphology due to these atomic steps, led to a drastic change in the electronic properties of the strained films. In **Figure 5,** the R(T) measurements reveal significant differences in the transport behaviour of $V_2O_3$ films depending on the sapphire substrate treatment. The film grown on unannealed sapphire exhibits a strongly suppressed MIT. In contrast, the film grown on sapphire annealed at 1000 °C shows a partial transition, with a resistance change of approximately one order of magnitude. Films grown on substrates annealed at 1200 °C display a fully developed MIT, with a resistance change of six orders of magnitude at the transition temperature 160 K. The negative *dR/dT* and high resistance near room temperature confirm the insulating phase of the film grown on sapphire annealed at 1400 °C. This film shows PI to AFI phase transition at 170 K.

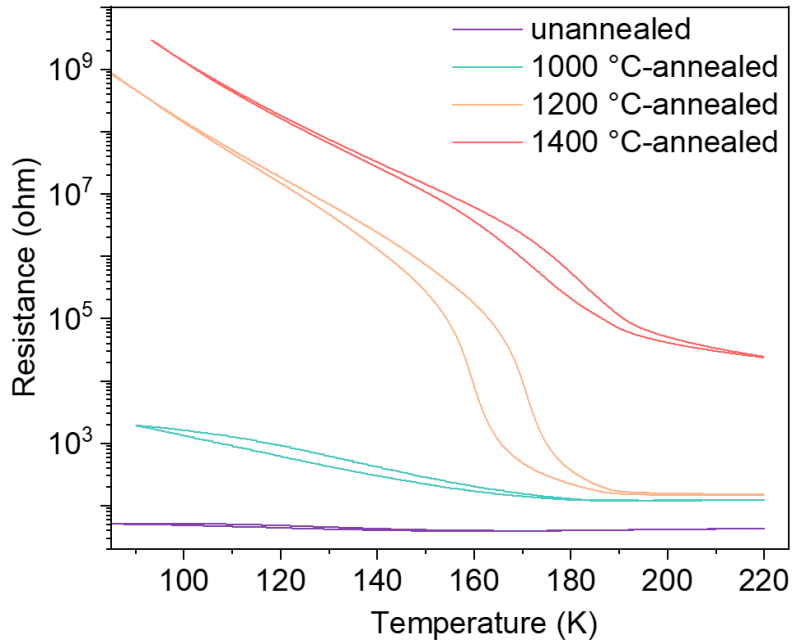

**Figure 5.** *Temperature-dependent resistance of $V_2O_3$ films grown on sapphire substrates annealed at different temperatures:* *The film on unannealed sapphire shows a suppressed MIT, while films on sapphire annealed at 1000 °C and 1200 °C display progressively stronger transitions, with the 1200 °C film exhibiting a six-order resistance change. At 1400 °C, a PI phase emerges at room temperature and accompanied by a PI-AFI transition.*

It is also noteworthy that the duration of annealing plays a crucial role in determining the dimensions of atomic steps on the substrate surface. As illustrated in **Supplementary figure 7 (a-c)**, annealing sapphire at 1000 °C for 6 hours results in relatively small atomic steps, whereas extending the annealing time to 24 hours leads to the formation of larger steps. These enlarged steps increase the likelihood of defect formation in the grown $V_2O_3$ films, which can facilitate strain relaxation and evolved MIT (**Supplementary figure 7 (d))**.

**DISCUSSION**

The STEM imaging and XRD analysis show the importance of the atomic steps induced by substrate annealing in promoting relaxation of compressive strain in the $V_2O_3$ films. However, this alone cannot explain the in-plane tensile strain observed in films grown on substrates annealed at 1400 °C. To explain this, we note that the films are grown at temperatures of ~1000 K, whereas $V_2O_3$ exhibits its rich phase diagram below 400 K. Therefore, we suggest that the thermal expansion mismatch between the sapphire and the $V_2O_3$ also plays a crucial role in determining the strain state. To examine this possibility, we performed temperature dependent XRD measurements of the $(11\bar{2}0)$ peak between 200-600 K (limited by our experimental setup), from which we extracted the a-parameter of the lattice (see **Supplementary figure 8**). We find that instead of following the expansion of the bulk $V_2O_3$ *a*-parameter, the *a*-parameter of the film expands similarly to that of sapphire. In **Figure 6(a)** we illustrate the effect of this clamping on the strain evolution in the $V_2O_3$ film: At the growth temperature the in-plane lattice parameters of the films will be between that of bulk $V_2O_3$ and sapphire depending on the extent of relaxation of the compressive in-plane strain. Upon cooling, the $V_2O_3$ *a*-parameter will contract more moderately than that of bulk $V_2O_3$ due to substrate clamping. If, at low T, the resulting film *a*-parameter is larger than that of bulk $V_2O_3$ the sample will enter a tensile strain regime. The key parameter which determines whether this occurs is the starting point of the strain at the growth temperature. With increasing relaxation of the film at the growth temperature, a more tensile strained state will form upon cooling. If compressive strain does not relax sufficiently during growth, the sample will remain with strong compressive strain, despite partial relaxation due to the thermal contraction mismatch. As found in this study, the key parameter which determines the relaxation is the substrate morphology, which facilitates the formation of dislocations and stacking faults in the vicinity of atomic steps **(Figure 6(b))**. By determining the strain state, the effect of the substrate morphology on the electronic properties of the $V_2O_3$ films is very substantial. In the most tensile strained films, the PI phase is stabilized at room temperature and the transition to the AFI occurs at ~180 K. In contrast, a nearly complete suppression of the transition is observed for the most compressively strained films, stabilizing the paramagnetic metallic phase at all temperatures.

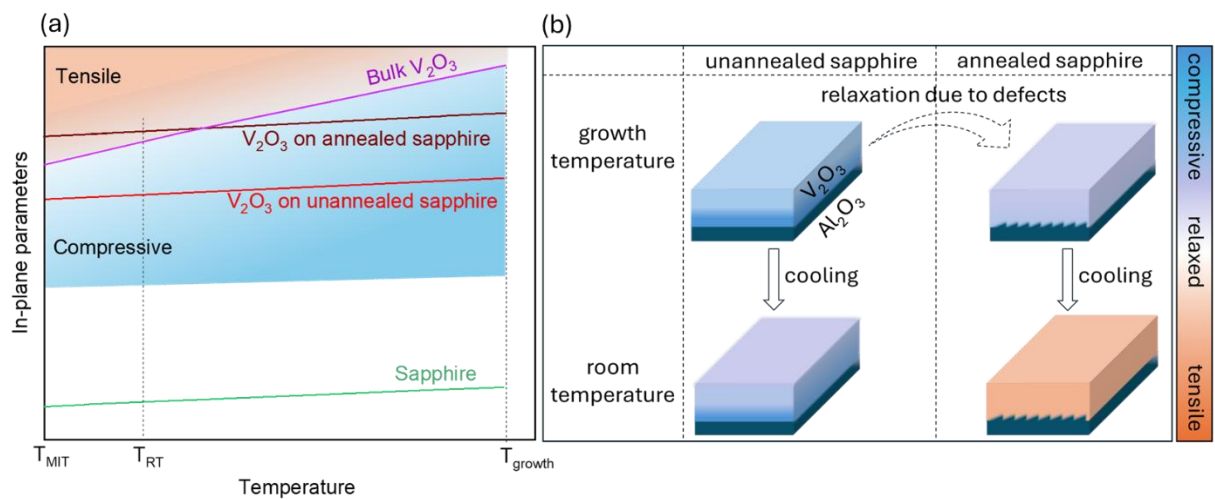

**Figure 6.** *Mechanism for formation of compressive/tensile in-plane strain in $V_2O_3$ films: (a) In-plane lattice parameters of bulk $V_2O_3$, $V_2O_3$ film, and sapphire as a function of temperature. At the growth temperature the in-plane lattice parameters of the films will range between those*

*of bulk V$_2$O$_3$ and sapphire depending on the extent of compressive strain relaxation, as determined by the substrate morphology. Due to clamping, the film in-plane lattice contraction (slope) follows that of the sapphire rather than that of bulk V$_2$O$_3$. Upon cooldown the film/bulk parameters may or may not intersect, resulting in tensile or compressive strain in the film at low T, respectively. (b) Schematics of the effect of the terrace morphology on the strain during growth and upon cooldown. For unannealed substrates large in-plane compressive strain develops in the film due to the lattice mismatch and partially relaxes due to thermal expansion mismatch. When grown on annealed sapphire, atomic steps in the substrate facilitate the formation of dislocations and stacking faults in the film which reduce the compressive strain at the growth temperature. After cooldown the thermal expansion mismatch produces in-plane tensile strain in the film.*

**CONCLUSIONS**

In summary, this study demonstrates that annealing sapphire substrates induces tuneable atomically stepped terraces, which play a pivotal role in modulating strain and transport properties in V$_2$O$_3$ thin films. High-resolution STEM data reveals that films grown on annealed sapphire exhibit structural defects that serve as relaxation pathways for compressive strain. Upon cooldown, clamping to the substrate reduces the compressive strain and may even result in tensile strain, depending on the extent of relaxation of the compressive strain during growth. Thus, minute changes in substrate morphology may induce reversal of the strain state in the films at low temperature, ranging from compressive to tensile states. Transport and XRD data confirm this scenario and show that controlling substrate morphology can induce drastic changes in film properties. These range from stabilization of the metallic phase at all temperatures (strong compressive strain), to fully developed metal-insulator transition (nearly relaxed) to fully insulating at all temperatures (strong tensile strain). Overall, atomic step-mediated strain engineering offers a powerful strategy to control strain and phase transitions and may be broadly applicable to other Mott materials. This strategy provides a pathway to tune electronic phase transitions for next-generation resistive switching applications.

**ACKNOWLEDGEMENTS**

This work was funded by the European Union's Horizon Europe research and innovation program (ERC, MOTTSWITCH, 101039986). Views and opinions expressed are however those of the authors only and do not necessarily reflect those of the European Union or the European Research Council Executive Agency. Neither the European Union nor the granting authority can be held responsible for them. This work is supported in part by a Technion fellowship.

# Reversal of strain state in a Mott insulator thin film by controlling substrate morphology

Reetendra Singh[1], Abhishek Rakshit[1], Galit Atiya[1], Michael Kalina[1], Yaron Kauffmann[1], and Yoav Kalcheim[1]*

[1]Department of Materials Science and Engineering, Technion-Israel Institute of Technology, Haifa 32000, Israel

*Email: ykalcheim@technion.ac.il

**Experimental methods**

Substrate morphology measurements:

*c*-plane sapphire substrates from MTI Corporation were annealed in air at 1000 °C, 1200 °C, and 1400 °C for 12 hours using a Kejia furnace. The morphology of the substrates was observed by Atomic Force Microscopy (AFM). The dimensions of atomic steps on the sapphire surface were determined from surface roughness measurements using Gwyddion software.

Growth of $V_2O_3$ thin films:

100 nm $V_2O_3$ films were grown on unannealed and annealed sapphire substrates of by RF magnetron sputtering in 5 mTorr Ar atmosphere and at a substrate temperature of 750 °C. Following growth, the samples were rapidly thermally quenched.

Structural characterization of the films:

To determine the lattice constants of the $V_2O_3$ films, we first aligned the symmetric (0006) and asymmetric ($20\bar{2}2$) reflections using a Rigaku Smartlab diffractometer equipped with a four-circle goniometer. The sample stage allows independent rotations about two perpendicular axes ($R_x$ and $R_y$), azimuthal rotation ($\varphi$), and tilt ($\chi$) in the diffraction plane. Measurements were performed using Cu K$\alpha_1$ radiation ($\lambda$ = 1.540593 Å). The (0006) peak, corresponding to the out-of-plane c-axis lattice spacing, was aligned in a symmetric $\omega$-$2\theta$ geometry. After coarse alignment on a strong substrate peak, we performed a high-resolution $\omega$-$2\theta$ scan around the nominal position of the (0006) reflection. The peak position was refined by adjusting sample height and tilt until maximum intensity was achieved.

For a hexagonal system, the interplanar spacing *d*:

$$\frac{1}{d^2} = \frac{4}{3}\left(\frac{h^2+hk+k^2}{a^2}\right) + \frac{l^2}{c^2} \tag{1}$$

The Bragg's law: $n\lambda = 2d \sin\theta$ (2)

Order of reflection (n= 1), $\lambda$ is the wavelength of X-rays (Cu K$_\alpha$= 1.5406 Å).

in-plane strain $\varepsilon_a$ (%) = $\left[\frac{a_{(film)} - a_{(bulk)}}{a_{(bulk)}}\right]$x100 (3)

Out-of-plane strain $\varepsilon_c$ (%) = $\left[\frac{c_{(film)} - c_{(bulk)}}{c_{(bulk)}}\right]$x100 (4)

Errors of in-plane and out-of-plane strains:

$\Delta\varepsilon_a$ (%) = $\left[\frac{\Delta a}{a_{(bulk)}}\right]$x100 (5)

$\Delta\varepsilon_c$ (%) = $\left[\frac{\Delta c}{c_{(bulk)}}\right]$x100 (6)

Using equations (1)-(2) and the measured (0006) and ($20\bar{2}2$) peak positions, the film's *c*- and *a*-axis lattice constants were determined. Strains and corresponding errors were calculated using equations (3)-(6) and shown in **Table 1** in the main text. The bulk $V_2O_3$ lattice parameters $a_{(bulk)}$ and $c_{(bulk)}$ were taken from [1] to calculate in-plane and out-of-plane strains and their errors.

Fits to XRR measurements were performed using the differential evolution algorithm GenX [2].

Transport measurements:

Electrical transport measurements were performed using a Lakeshore TTPX probe station equipped with a Keithley 2450 source meter unit. Probes were placed on indium contacts prepared on the $V_2O_3$ samples.

HRSTEM:

Lamellas were fabricated along the $[\bar{1}010]$ direction of sapphire from a 100 nm thick $V_2O_3$ film using a Thermo Fisher Helios 5 Plasma Focused Ion Beam (PFIB).

High-resolution imaging was performed on a Thermo Fisher/FEI Titan Themis TEM, operated at 300 keV and using the HAADF-STEM detector.

**Supplementary figures**

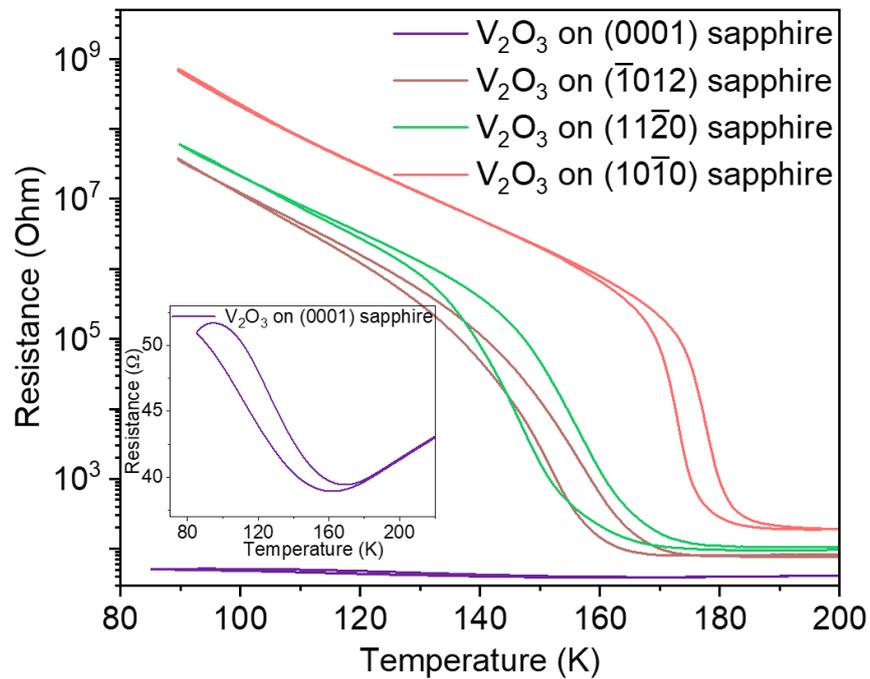

**Supplementary figure 1.** Resistance versus temperature for a 100 nm $V_2O_3$ film grown on sapphire substrates with different orientations. Films on *c*-sapphire shows a strongly suppressed metal-insulator transition, whereas films grown on other sapphire orientations exhibit a full transition of 5-7 orders of magnitude. The inset presents the resistance–temperature curve of $V_2O_3$ films grown on (0001)-sapphire under similar. Under substrate-induced 2D confinement, $V_2O_3$ film on (0001)-sapphire faces restricted in-plane (*ab*-plane) expansion, necessary to go to the monoclinic phase. In other orientations, out-of-plane expansion eases this distortion.

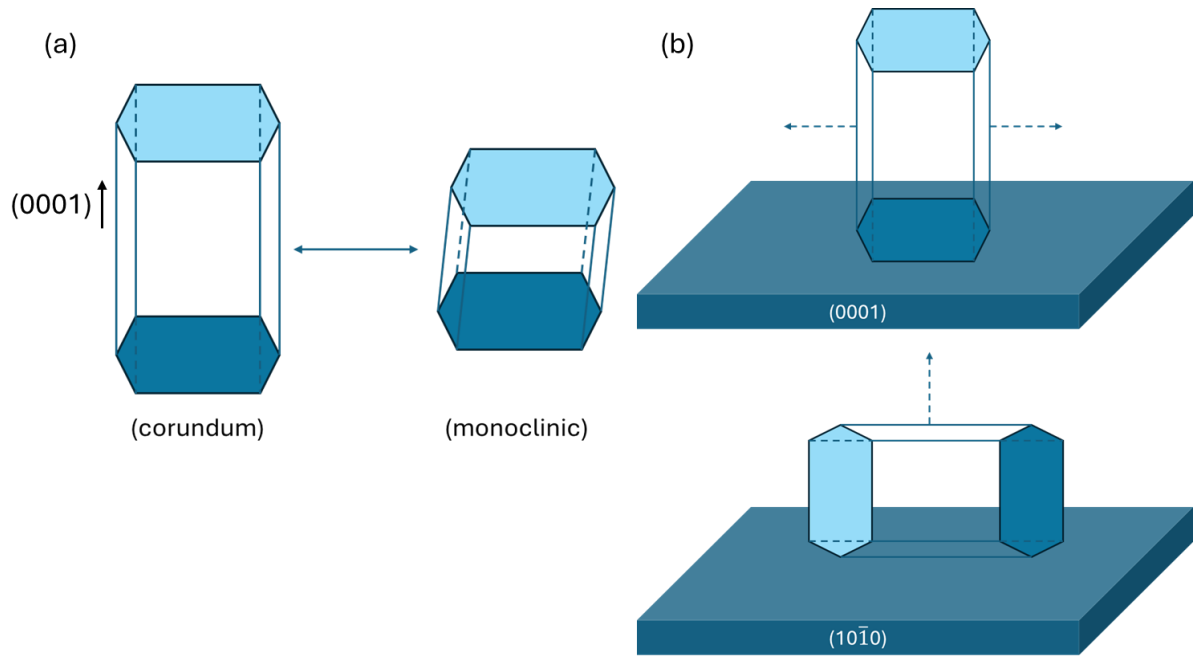

**Supplementary figure 2.** (a) Graphical illustration that during the MIT, the monoclinic distortion causes the [0001] direction to contract while the perpendicular directions expand. (b) In (0001)-oriented growth, the lattice expansion is restricted to the film plane, so the monoclinic transition increases in-plane compressive stress; in contrast, other orientations allow out-of-plane expansion, leading to minimal strain from the monoclinic distortion.

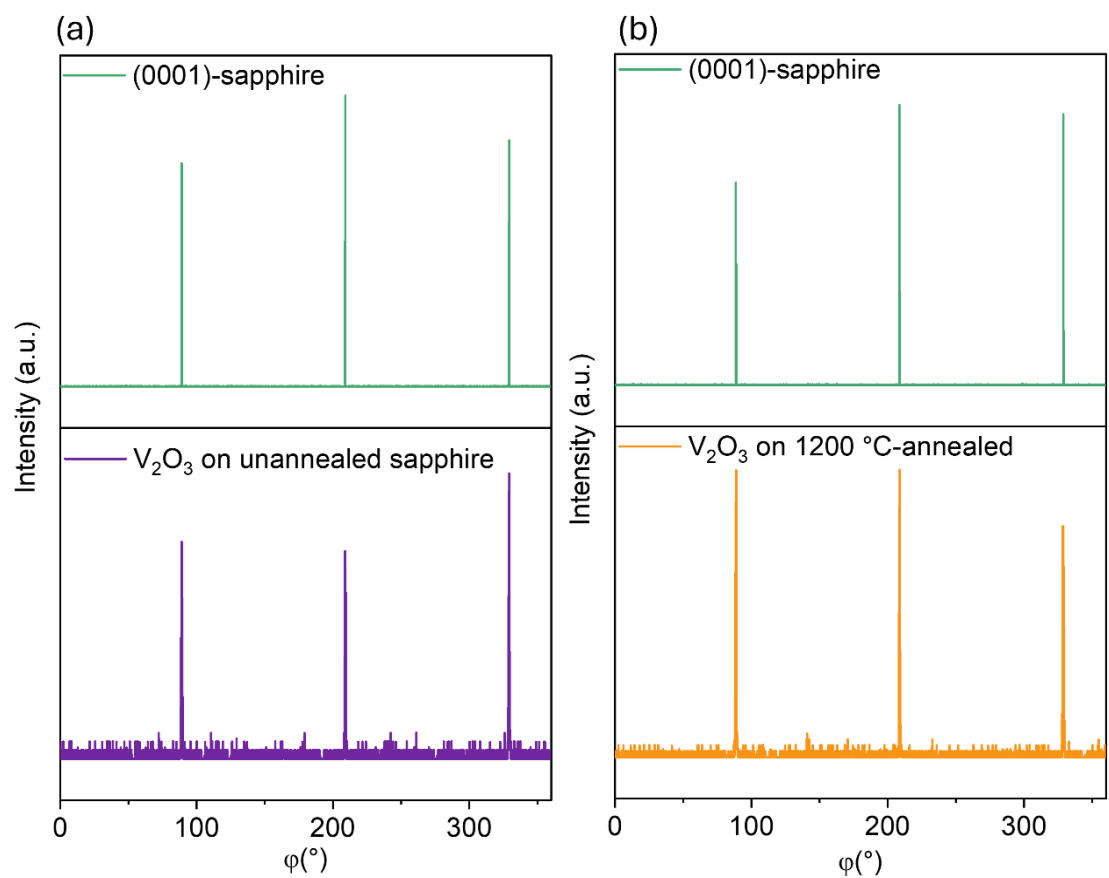

**Supplementary figure 3.** $\phi$-scans for the $(20\bar{2}2)$ peak of (a) sapphire and V$_2$O$_3$ on unannealed sapphire and (b) V$_2$O$_3$ on annealed sapphire at 1200°C showed three peaks at intervals of 120°, following the crystallographic orientation of the sapphire, revealing an epitaxial relationship between the films and substrates.

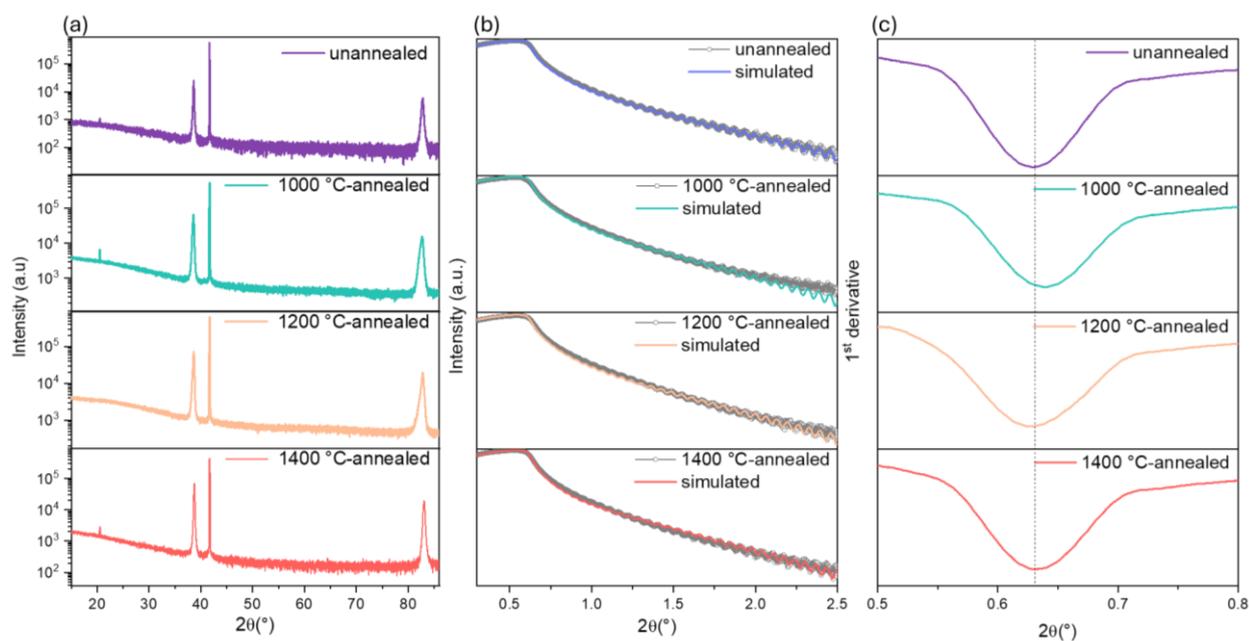

**Supplementary figure 4.** (a) XRD patterns of $V_2O_3$ films grown on *c*-plane sapphire substrates annealed under different conditions, showing no noticeable change in film properties, confirming structural consistency across all samples, (b) XRR and (c) its first derivative for $V_2O_3$ films grown on *c*-plane sapphire with varying annealing conditions, confirming similar film quality, thickness and density across all samples.

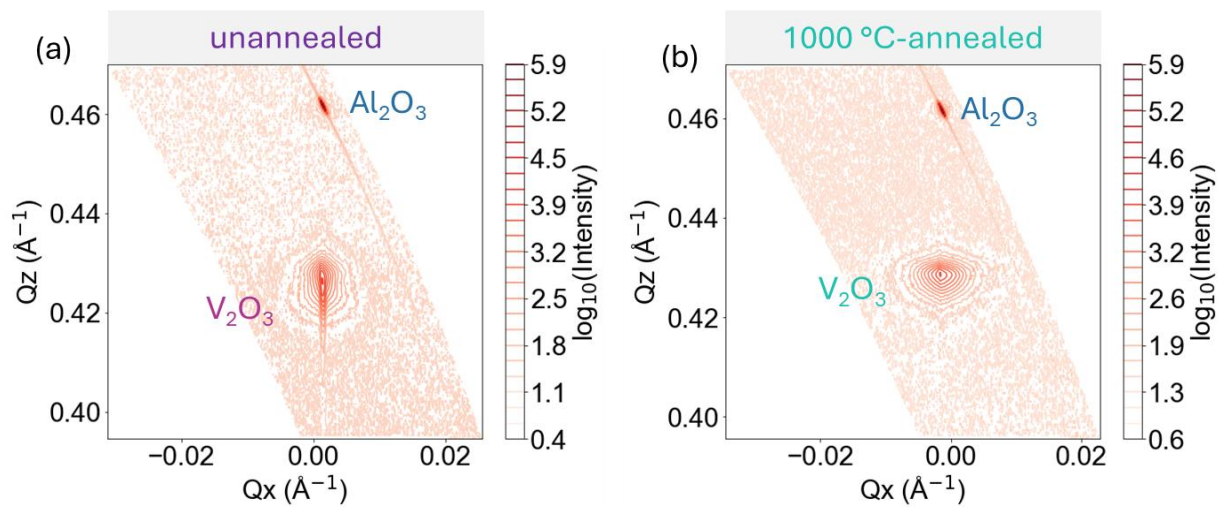

**Supplementary figure 5.** The RSM of the (0006) reflection for films grown on (a) unannealed sapphire and (b) annealed sapphire at 1000°C shows that films on unannealed sapphire exhibit a strain gradient along the c-axis due to substrate clamping. This strain is relieved in the films grown on annealed sapphire, resulting in film tilting and disordering.

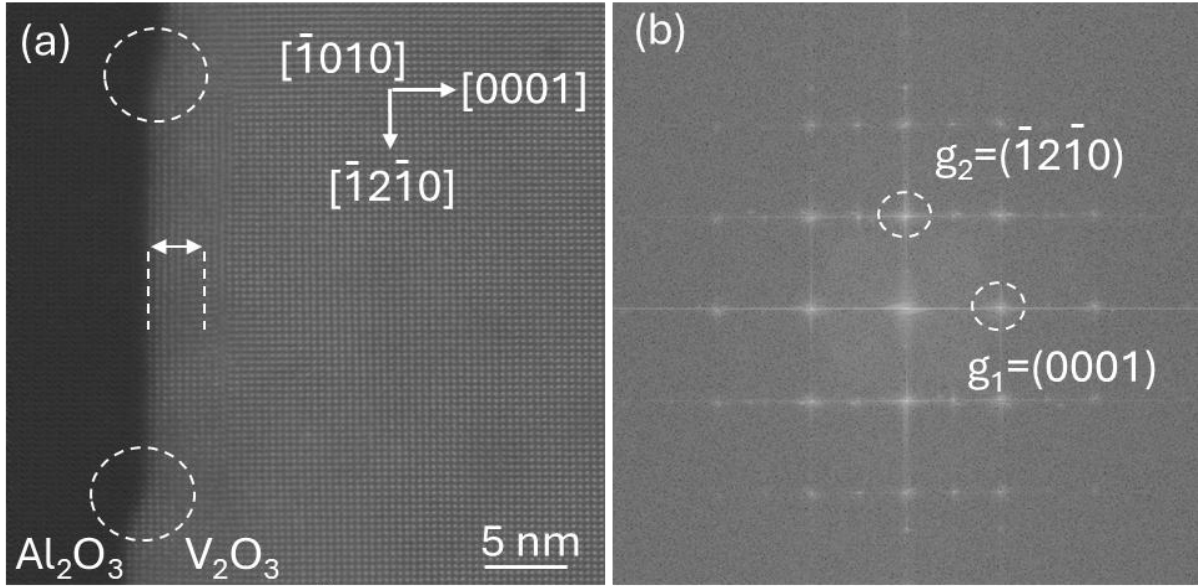

**Supplementary figure 6**. (a) HRSTEM image showing two adjacent atomic steps on sapphire annealed at 1200 °C (see sotted circles). (b) The Fourier-transform of the of the lattice and g-vectors chosen for GPA.

Geometric phase analysis (GPA) was carried out using the *Strain ++* open-source software package [3]. The two intense reflections in the power spectrum, $g_1=(0001)$ and $g_2=(\bar{1}2\bar{1}0)$, were selected to compute the displacement field, enabling the generation of local strain maps and filtered Bragg images for defect analysis.

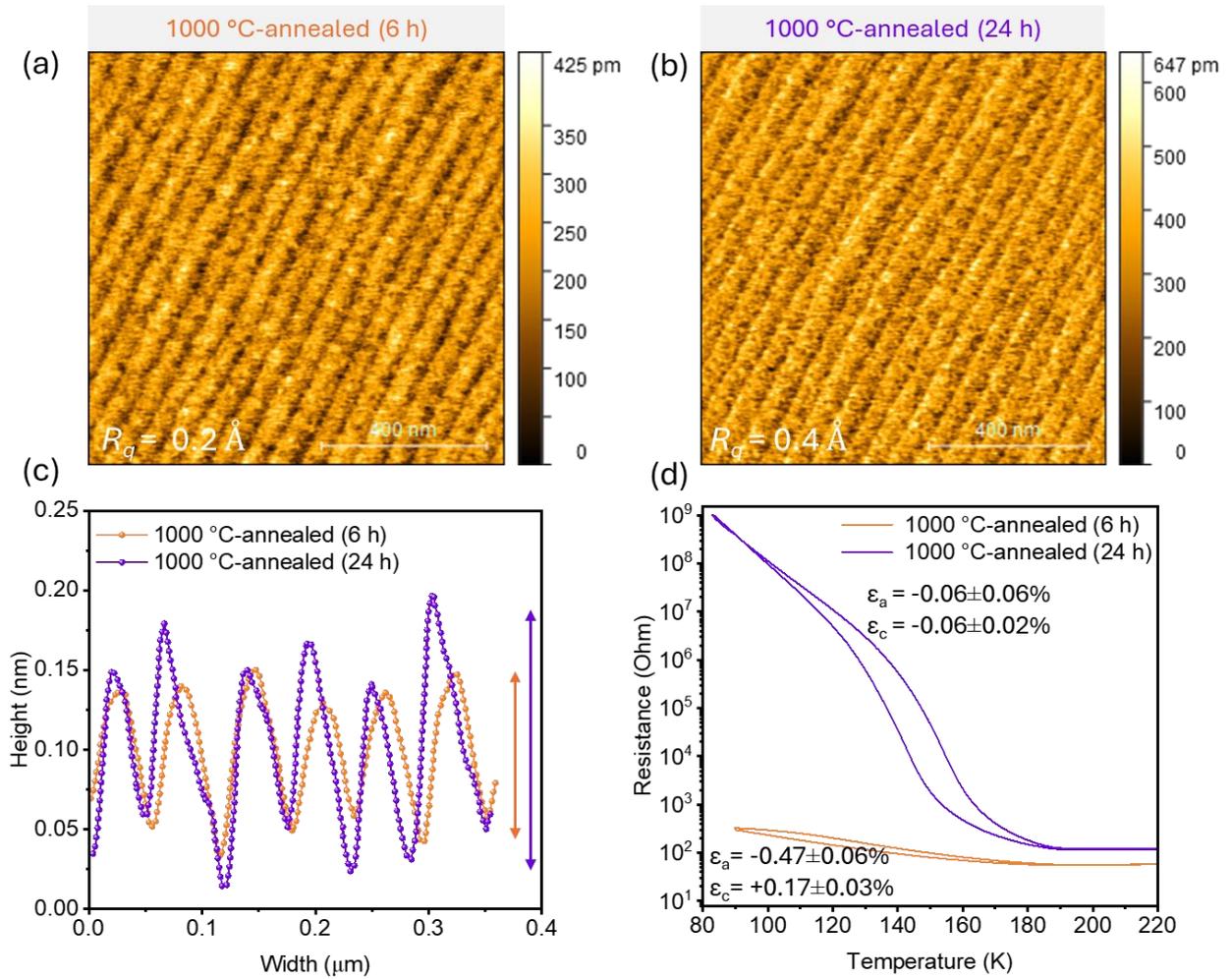

**Supplementary figure 7.** (a, b) AFM images of annealed *c*-plane sapphire showing atomic steps; (c) corresponding line profiles indicating the step height and width for substrates annealed at 1000 °C for 6 h and 24 h, respectively; (d) temperature-dependent resistance of $V_2O_3$ films grown on these annealed sapphire substrates.

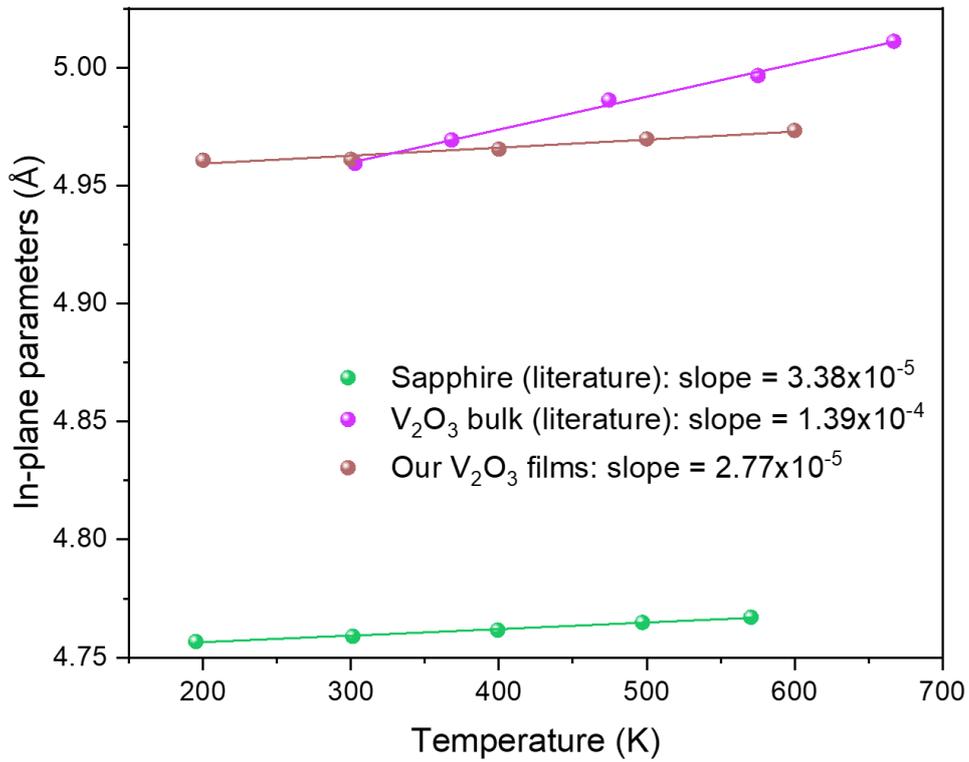

**Supplementary figure 8.** In-plane lattice parameters of bulk $V_2O_3$, $V_2O_3$ film, and sapphire as a function of temperature. At the growth temperature the in-plane lattice parameters of the films will range between those of bulk $V_2O_3$ and sapphire depending on the extent of compressive strain relaxation, as determined by the substrate morphology. Due to clamping, the film in-plane lattice contraction follows that of the sapphire rather than that of bulk $V_2O_3$. Upon cooldown the film/bulk parameters may or may not intersect, resulting in tensile or compressive strain in the film at low T, respectively. The *a*-parameters for sapphire and bulk $V_2O_3$ were taken from the previous reports [4-6].

**Supplementary references**